\documentclass[page-classic]{epl2} 
\usepackage{graphicx}

\newcommand{\bfa}{{\cal W}\!\!\!\!\!\!{\cal W}}

\title{Gauge symmetry breaking and topological quantization for the Pauli Hamiltonian}

\author{Ernesto Medina\inst{1,2}\and Alexander L\'opez\inst{1}\and Bertrand Berche\inst{2} }
\institute{
\inst{1} Centro de F\'\i sica, Instituto Venezolano de Investigaciones Cient\'\i ficas, Apartado 21874, Caracas 1020-A, Venezuela\\
\inst{2} Laboratoire de Physique des Mat\'eriaux, Universit\'e Henri Poincar\'e, Nancy 1, 54506\\
Vand\oe uvre les Nancy Cedex, France}


\abstract{
We discuss the Pauli Hamiltonian within a ${SU(2)}$ gauge theory interpretation, where the gauge symmetry is broken.  This interpretation carries directly over to the structural inversion asymmetric spin-orbit interactions in semiconductors and offers new insight into the problem of spin currents in the condensed matter environment. The central results is that symmetry breaking leads to zero spin conductivity in contrast to predictions of Gauge symmetric treatments. Computing the translation operator commutation relations comprising the simplest possible structural inversion asymmetry due to an external electric field, we derive a new condition for orbit quantization. The relation between the topological nature of this effect is consistent with our non-Abelian gauge symmetry breaking scenario.}

\begin{document}

\maketitle

The role of spin-orbit interaction in condensed matter systems, be it of intrinsic or extrinsic sources, is of great interest due to its potential role in the generation of spin polarized currents, spin filters, spin accumulation\cite{SarmaReview}, spin optics\cite{BalseiroUsaj} and in general the quantum anomalous spin Hall effect in the absence of magnetic fields. Two generic kinds of systems have been studied for the possibility of spin currents: the first kind consists of partially filled bands (metallic) in three dimensional p-doped semiconductors\cite{NagaosaScience} and two dimensional electron systems with Rashba type couplings\cite{Sinova}. Such systems exhibit spin currents that are specially sensible to disorder effects depending on the relative values of the spin orbit energy scale, and the energy scale $\hbar/\tau$ associated with the scattering time\cite{SchliemannLoss}. Other works have assessed that any weak disorder may destroy the Spin Hall Effect in the two dimensional Rashba model for infinite samples\cite{Halperin}\cite{Molenkamp}.
The second kind of systems are topologically protected by band gap insulators, as in the integer quantum Hall effect\cite{KaneMele}\cite{NagaosaInsulator}, which are robust against disorder as long as the gap is not destroyed or very small\cite{Min}.

Sheng et al\cite{Haldane2} recently established a relation between the IQSHE and topological Chern numbers characterizing the bulk states and the existence of edge states in graphene\cite{Novoselov}. Although this scenario is promising, the value of the gap opened up by the SO interaction in graphene is quite small ($10^{-3} {\rm meV}$) so disorder is important. On the other hand, work by Murakami, Nagaosa and Zhang\cite{NagaosaInsulator} studies a class of zero gap and narrow gap band insulators. The model assumes time reversal or space inversion symmetries and doubly degenerate parabolic bands close to the Fermi energy. Here the gap protecting against disorder is also generated by the spin-orbit coupling while the system is three dimensional. A restriction to two dimensions of the model in ref.\cite{NagaosaInsulator} is qualitatively different, as shown by Bernevig and Zhang\cite{Bernevig} and Jin, Li and Zhang \cite{ZhangYangMills} where the SHE is found to be quantized. A recent experimentally feasible proposal by Bernevig, Hughes and Zhang\cite{ZhangScience} for a topological phase transition in HgTe quantum wells shows a Quantum Spin Hall phase.

In the present paper we propose a novel scenario for Quantized spin transport in a two dimensional system . We follow the approach of formulating the Pauli Hamiltonian in terms of an $SU(2)$ theory with the corresponding generalized vector potential, acting on spin currents\cite{ZhangYangMills,Tokatly}. Our reformulation of previous approaches brings about important physical features where the gauge symmetry of the Yang-Mills vector potential is obviously broken. This has multiple physical implications in the Yang-Mills context, i) The vector potential is a physically observable quantity, in this case related to the electric field or crystal potential: ii) The vector potential exerts a torque on the spin and makes for a definite separation between matter and radiative contributions which are gauge dependent in the symmetric Yang-Mills theory. iii) There is the issue of the topological nature of closed paths: phases of open and closed paths are physical when the gauge symmetry is broken so only the so called `local' topological effects are possible (phases are produced by physical fields)\cite{AharonovReznik,PeshkinLipkin}. We will discuss more on these ideas below.

Neglecting electron-electron interactions, we start by considering the Pauli Hamiltonian to order $v/c$
\begin{eqnarray}
H=\frac{\left ({\mathbf p}-e{\mathbf A}\right)^2}{2m_0}+V+\frac{e\hbar}{2m_0}{\boldsymbol \sigma}\cdot {\mathbf B}-\frac{e\hbar{\boldsymbol \sigma}\cdot \left ({\mathbf p}-e{\mathbf A}\right)\times{\bf E}}{4m_0^2c^2}
\mbox{}-\frac{e\hbar^2}{8m_0^2c^2}\nabla\cdot{\bf E},\label{PauliFirst}
\end{eqnarray}
where the first and second terms are the kinetic energy, including the minimal coupling to the electromagnetic field, and the substrate potential denoted by $V$, that can be assumed periodic while $m_0$ is the bare mass of the electron. The third term is the Zeeman term where $\mathbf B$ is the magnetic field, $e$ is the electron charge and $\boldsymbol \sigma$ is the Pauli matrix vector. The fourth and fifth terms are the spin-orbit interaction and the Darwin terms, where $\bf E$ is the electric field and $c$ the speed of light. We have assumed a static potential so that the rotor of the electric field is absent.  In order to suggest an $SU(2)\times U(1)$ form we can rewrite the Hamiltonian, following Jin, Li and Zhang\cite{ZhangYangMills} as
\begin{eqnarray}
\label{HLambda}
H=\frac{1}{2m_0}\left ({\mathbf p}-e{\mathbf A}-\frac{2m_0}{\hbar^2}{\mathbf s}\times {\boldsymbol \lambda}\right )^2
+\frac{1}{2}{\boldsymbol \nabla}\cdot {\boldsymbol \lambda}-\frac{2}{\hbar}\mu_B{\boldsymbol s}\cdot {\mathbf B} -\frac{m_0\lambda^2}{\hbar^2}+V,\label{PauliSecond}
\end{eqnarray}
where ${\boldsymbol \lambda}=-\frac{e\hbar^2}{4m_0^2c^2}{\bf E}$ and ${\mathbf s}=\hbar{\boldsymbol\sigma}/2$ for spin $1/2$. The terms involving ${\boldsymbol \lambda}$ can also contain ${\boldsymbol \nabla} V$, nevertheless this term produces no contribution to the spin splitting if the potential is periodic and inversion-symmetric. As noted in ref.\cite{ZhangYangMills}, one can identify the third term in the square as a new $SU(2)$ connection defined by $g{W}_i^a\tau^a=-(e\hbar/2 m_0c^2)\varepsilon_{iaj}{ E}_j\tau^a$, where the $\tau^a$ are the symmetry generators for $SU(2)$ obeying the commutation relation $[\tau^a,\tau^b]=i\varepsilon_{abc}\tau^c$, $\varepsilon_{abc}$ being the totally antisymmetric  tensor and $g=\hbar$.  The relation between the spin operator and the corresponding generators is $\hbar\tau^a=s^a$ and the spin is $s=1/2$. Substituting these definitions we can rewrite Eq.\ref{PauliSecond} as
\begin{eqnarray}
H=\frac{1}{2m_0}\left ({\mathbf p}-e{\mathbf A}-g{\bf W}^a\tau^a\right )^2-\frac{g^2}{8m_0}{W}_i^b{W}_i^b
+g{W}_0^a\tau^a+e{A}_0,\label{PauliGauge}
\end{eqnarray}
where $g{W}_0^a\tau^a=-(e/m_0){\mathbf s}\cdot{\mathbf B}$, following the usual convention for a quadri-vector ${\bfa}^a= (W_0^a/c,W_i^a)$  and $A_0=V/e$. The latin indices from the middle of the alphabet, $ijk$, denote the spatial coordinates and run over three values, and the letters {\it a,b,c} denote the internal space indices for the generators. We will always use the internal space indices as superindices. This formulation differs from that of ref.\cite{ZhangYangMills,Tokatly} most importantly in the second term, where it is taken a part of the scalar potential. The purpose of writing the second term here, as a function of the $SU(2)$ connection is to evidence {\it gauge symmetry breaking} (GSB) in this Hamiltonian. This observation has important consequences in the physical interpretation of the resulting Yang-Mills fields and is the reason why the Yang-Mills fields themselves are observable quantities, where as in a gauge symmetric theory they would be gauge dependent\cite{weinberg}. 

In order to generate the Noether currents in a canonical fashion, one must formulate the appropriate Lagrangian for the corresponding equations of motion. The non-relativistic Lagrangian density we seek is 
\begin{eqnarray}
{\mathcal L}= \frac{i\hbar}{2}\left (\Psi^{\dagger}\dot{\Psi}-\dot{\Psi}^{\dagger}\Psi\right )&-&\Psi^{\dagger}\left (\frac{-g^2}{8m_0}{W}_i^b{W}_i^b+g{W}_0^a\tau^a+e{A}_0\right)\Psi-\frac{\hbar^2}{2m_0}\left [ {\mathcal D}_j\Psi\right ]^{\dagger}{\mathcal D}_j\Psi\nonumber\\
&-&\frac{e^2}{4m_0}F_{\mu\nu}F_{\mu\nu}-\frac{g^2}{4m_0}{G}^a_{\mu\nu}{ G}^{a}_{\mu\nu},
\end{eqnarray}
where $\Psi$ is a Pauli spinor, ${G}_{\mu\nu}^a =\partial_\mu{W}_\nu^a-\partial_\nu{W}_\mu^a
 -\epsilon^{abc}{W}_\nu^b{W}_\mu^c$ and ${F}_{\mu\nu} =\partial_\mu A_\nu-\partial_\nu A_\mu$ are the $SU(2)$ and $U(1)$ field tensors respectively. The new term $\frac{-g^2}{8m_0}{W}_i^b{W}_i^b$ is due to gauge symmetry breaking. The covariant derivative is then of the form ${\mathcal D}_i=\partial_i-\frac{ie}{\hbar}A_i-\frac{i g}{\hbar}{W}^a_i\tau^a$. This form of the covariant derivative determines the well known $U(1)$ coupling constant $e/\hbar$ and the $SU(2)$ coupling constant for this theory is $g/\hbar$. The Hamiltonian in Eq.~\ref{PauliGauge} is derived from the corresponding Lagrange equations for the matter fields $\Psi$. We can now derive the conserved current $\mathcal J^a_{\nu}$ in the ordinary sense, from the equations of motion\cite{weinberg}. This is the full spin current carried both by matter and radiation i.e. ${\cal J}\!\!\!\!\!{\mathcal J}^a={\boldsymbol J}^a_M+{\boldsymbol J}^a_R$. This full current is gauge invariant if the Lagrangian is gauge invariant. The spatial (spin current) and time (magnetization) components of the current density then follow as
\begin{eqnarray}
\label{totalspatialcurrent}
{\mathcal J}_i^a=\Psi^{\dagger} \left ( \frac{g^2}{4m_0}{W}^a_i\right )\Psi+\frac{g^2}{m_0}\varepsilon^{abc}{W}^b_\nu{G}^{c}_{\nu i}
-\frac{i\hbar g}{2m_0}\left [ \left (\tau^a\Psi\right)^{\dagger}{\mathcal D}_i\Psi
-\left ({\mathcal D}_i\Psi\right )^{\dagger}\left (\tau^a\Psi\right ) \right ],
\end{eqnarray}
and the spin polarization
${\cal J}^a_0=\Psi^{\dagger}g\tau^a\Psi+\frac{g^2}{m_0}\varepsilon^{abc}{W}^b_j{G}^{c}_{j0}$.
Three terms can be distinguished in the spin current; the third term has the canonical form expected for the material current namely $(J_M)^a_i=(g/2)\Psi^{\dagger}(\tau^a v_i+v_i\tau^a)\Psi$, where $v_i=(1/i\hbar)[r_i,H]$. The second term is the canonical radiative contribution originating from the derivative with respect to the gauge potential of the non-Abelian contribution of the field tensor ${G}^{a}_{\mu\nu}$. Finally, the first term comes from the gauge symmetry breaking contribution. The last two terms were already described in ref.\cite{ZhangYangMills} as taken from an apparently gauge symmetric form. The magnetization term has both a material contribution (the first term) and a radiative contribution as both matter and radiation carry angular momentum. We emphasize that, the extent to which gauge symmetry is broken depends on the choice of the electric field. If only the $E_z\ne 0$, then one allows gauge transformations that leave $W_1^2=-W_2^1$ invariant. This is analogous to the remnant $Z_2$ group after $U(1)$ GSB in superconductors and a similar situation in the electro-weak GSB mechanism\cite{weinberg}.

Rewriting the last term in Eq.\ref{totalspatialcurrent} in terms of ordinary derivatives plus a gauge dependent term we obtain
\begin{equation}
-\frac{i\hbar g}{2m_0}\left [ \left (\tau^a\Psi\right)^{\dagger}{\partial}_i\Psi
-\left ({\partial}_i\Psi\right )^{\dagger}\left (\tau^a\Psi\right ) \right ]-\Psi^{\dagger}\left (\frac{g^2}{4m_0}W_i^a\right )\Psi.
\end{equation}
The second term is the non-Abelian analog of the London term in superconductivity. The main result of this paper is then to recognize that such second term exactly cancels the symmetry breaking term in Eq.\ref{totalspatialcurrent} and renders {\it zero} matter currents proportional to the electric field (zero spin conductivity). The scenario is now very different from superconductivity: there, the London term is the only one remaining after symmetry breaking, while for the non-Abelian case, the London contribution gets cancelled. As discussed in references \cite{Tokatly,RashbaEquil}, equilibrium currents remain in relation to the leftover radiative contribution, cubic in the non-Abelian potential plus a field independent matter contribution.

In contrast to $U(1)$ gauge theory, the $SU(2)$ theory has new subtleties in relation to gauge symmetry: In $U(1)$ theory the electromagnetic field tensor is both gauge covariant and gauge invariant, due to the Abelian nature of the theory. From this, one derives the fact that the electric and magnetic fields remain unchanged by the gauge choice. This is also true of the charge current which is gauge invariant. In $U(1)$ the material current carries all the charge current since the radiative part is chargeless. The material current is thus locally conserved.  In the $SU(2)$ theory both the material and radiative currents carry angular momentum, so it is the total current that is related to any changes in the local angular momentum density. This is made obvious by writing the $SU(2)$ covariant non-Abelian Maxwell equation ${\mathcal D}_\mu {G}^{a}_{\mu\nu}=(J_M)_{\nu}^{a}$\cite{ZhangYangMills} (gauge symmetric situation). There is a torque on the material current carrying angular momentum due to the radiative part. In principle, any initial material current can be degraded towards the radiative component\cite{Sinova}. On the other hand, the torque may also be seen  to exert a radiation field that has served as a basic tool for angular momentum transfer in ferromagnetic multilayers\cite{MagneticCollection}.
This situation changes when the gauge symmetry is broken as happens here; The broken symmetry tells us how much matter and radiative components there are, and this separation is no longer gauge dependent because the gauge has been chosen.

Once the symmetry broken theory is established we can now determine the topological consequences of an electric field induced spin-orbit interaction in the presence of a periodic potential. In the following
we derive the commutator of `magnetic' operators for $SU(2)$ position independent gauge potential, that already produces interesting topological effects for the spin\cite{NagaosaScience}. We use as guide, the integer Hall case explained in detail in ref.\cite{Kohmoto}. The position dependent case leads to further complications due to the non-commutation of the rotation generators and will be treated elsewhere. For two dimensions and in the absence of the {\it U(1)} gauge potentials, we define the primitive translation vectors for the lattice as ${\bf v}$ and ${\bf w}$, so an arbitrary lattice point is described by ${\bf R}=n{\bf v}+m{\bf w}$. The `magnetic'  translation operators in the absence of $U(1)$ fields is
\begin{equation}
T_{\bf R}=\exp\left (i{\bf R}\cdot\left({\bf p}-g{\bf W}^a\tau^a\right) \right ).
\end{equation}
Such operators are {\it bona fide} translation operators since they commute with the Hamiltonian including the potential $V$, periodic in the lattice vectors ${\bf w}$ and $\bf v$. Nevertheless, commutators of different translations need not be zero due to the nontrivial commutation of the Pauli matrices. Consequently, we must see the conditions required for the operators to commute. Since here we have that $[p_i,W_i^a]=0$ for constant external electric fields, we write
\begin{eqnarray*}
[T_v,T_w]=\exp(i{\bf p\cdot(v+w)})
[\exp(-i\frac{g}{\hbar}{\bf v}\cdot{\bf W}^a\tau^a),\exp(-i\frac{g}{\hbar}{\bf w}\cdot{\bf W}^b\tau^b)].
\end{eqnarray*}
Using the identity $\exp(\pm i g\theta \sigma_n)=\cos g\theta\pm i\sigma_n\sin(g\theta)$, the expression for the commutator takes the form
\begin{equation}
\label{commutatorwide}
[T_v,T_w]=-4i\exp(i{\bf p\cdot(v+w)})\frac{\sin\left (\frac{2m_0}{\hbar^2}|{\bf v\times \boldsymbol{\lambda}}|\right)}{|{\bf v\times \boldsymbol{\lambda}}|}\frac{\sin\left (\frac{2m_0}{\hbar^2}|{\bf w\times \boldsymbol{\lambda}}|\right )}{|{\bf w\times \boldsymbol{\lambda}}|}{\boldsymbol{\lambda}\cdot (\bf v\times w}) {\boldsymbol \lambda}\cdot{\boldsymbol \tau}.
\end{equation}

\begin{figure}
\includegraphics[width=10 cm]{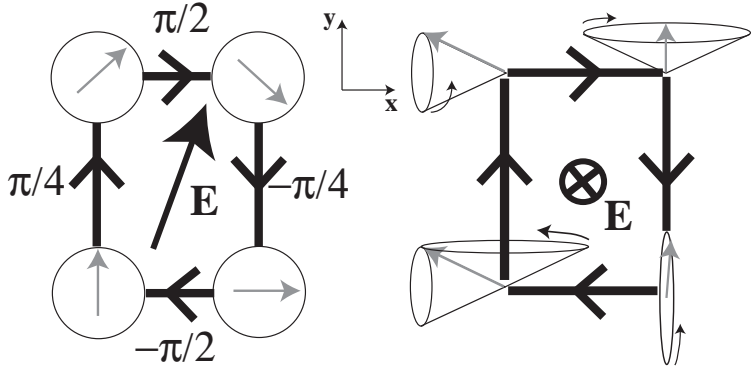}
\caption{Electric field a) in the plane permits closure for all precession paths no matter the field strength, what is done in one leg is undone by the other due to the spatial independence of the vector potential and the commutativity of rotations in two dimensions. b) When the electric field is perpendicular to the plane, uniqueness of the wave function occurs only fine tuning the electric field or defining a precession Brillouin zone. The right panel shows a generic spin vector director precessing first about the $x$ axis and integer number of cycles, and then about the $y$ axis by an arbitrary angle so that we can close the loop, rendering the wavefunction unique.}
\label{fig1}
\end{figure}
We can directly read the conditions for the translation operators to form a complete set of commuting operators from the previous expression. First, if the electric field (through ${\boldsymbol \lambda}$) is in the plane of the crystal vectors, the lattice translation operators always commute and no special conditions arise that quantize the closed trajectories. This fact is shown in terms of the precession of the spin-vector in Fig.\ref{fig1}. For electric fields in the plane the precession axis is always perpendicular to the plane, and all precession occurs as finite rotations in two dimensions. This renders the problem Abelian with a vector potential independent of space i.e. all trajectories close no matter the strength of the field or the lengths of the arms involved in the loop. Note that this conclusion is also independent of the spin direction. 

Non-trivial zeroes come from a field perpendicular to the crystalline plane, through the sinusoidal terms. Such terms require that 
$\omega=(2m_0/\hbar^2)|{\bf v}\times{\boldsymbol\lambda}|=\pi p$,
where $p$ is an integer. This relation is analogous to the quantization of the magnetic flux in the $U(1)$ integer quantum Hall problem. The electric field times the length of one dimension of the Brillouin zone must be a multiple of $2\pi m_0c^2/e$, a quantum for an open electric path integral i.e. an electric Brillouin zone. If we take an electric field such that $\omega=2\pi p/q$ (with $p/q$ a rational number) then we should choose a spin-precession Brillouin zone (analogous to the magnetic Brillouin zone in the $U(1)$ case or integer charge quantum Hall regime), $q$ times larger in the $\bf v$ direction in order for the translation operators to commute, or in other words, for there to be an integer number of cycles of the spinor along the trajectory. The closed path conditions securing uniqueness of the wave function is actually reduced to constraining just one direction in the Brillouin zone. 

The ultimate source of the non-Abelian character of the Yang-Mills theory is that three dimensional rotations generically do not commute. How paths close to render the wavefunction single valued is actually then an exercise in finding the condition under which three dimensional rotations commute. In Fig.\ref{fig1}, on the right panel, we show a generic spin direction made to precess, first about  the negative $x$ axis, around which the spin performs a complete number of cycles. Then there is a precession around the positive $y$ axis, through some arbitrary angle (in the figure $\pi/2$ for ease). We then close the circuit by first precessing around the positive $x$ axis (the change in direction of motion inverts the rotation axis) for a complete number of rotations and then around the negative $y$ axis by the same arbitrary angle as the second leg. This closes the path with the only condition that for one or both of the legs, the precession completes a full number of cycles as indicated by Eq.\ref{commutatorwide}.

The canonical form of computing phase changes on a path, be it $U(1)$ or $SU(2)$ theories is the Wilson loop\cite{weinberg}.  As the gauge symmetry is broken, we can define it on an open path along the edge, say $\bf v$, of the Brillouin zone. Taking the diagonalized form of the Wilson loop one obtains
\begin{eqnarray}
{\rm Diag}\left [\frac{2}{\pi}\int_0^v W_i^a\tau^a dx_i\right ]={\rm Diag}\left [\frac{e\tau^a}{\pi m_0c^2}\int_{0}^v \left ({\bf E}\times d{\bf x}\right )^a\right ]
=p\sigma^z,
\end{eqnarray}
where we have used the quantization condition. The integer $p$ plays the role of the Chern number for the $U(1)$ case. The Pauli matrix appears because, in the quantization condition, $p$ appears in pairs if one integer appears its negative is also a good quantum number.

We have argued that an open trajectory is physically meaningful here i.e. gauge invariant, since the gauge symmetry is broken as discussed above. This is not possible in the gauge symmetric $U(1)$ case since only closed trajectories are gauge invariant. This also relates to the recent discussion in the literature on the distinction between the Aharonov-Bohm effect and the Aharonov-Casher\cite{PeshkinLipkin,AharonovReznik,Spavieri}. In the Aharonov-Bohm effect there is no physical field (it is the vector potential) coupled to the change in the phase of the wave function, so there is no meaning to the local change in phase, while in the Aharonov-Casher effect the local phase change is meaningful, related to a real local contemporary field at any particular location. Such a physical distinction here is relevant, but both cases produce a discrete set of instances where the translation operators themselves commute.

The spin-orbit interaction is very small owing to the relativistic origin of the bare spin-orbit term in the Pauli Hamiltonian, that is inversely proportional to the rest mass energy of the electron. Any observable spin-orbit effect would then originate from very strong electric fields. Nevertheless for semiconductors, within the framework of ${\bf k}\cdot{\bf p}$ theory, we have the same form of the Pauli Hamiltonian\cite{Winkler}. The spin-orbit interaction depends on the smaller semiconductor gap and intrinsic spin-orbit gap of the bulk band structure, rendering a much larger effect. The Rashba term for a narrow gap semiconductor is of the form $\alpha(\sigma_z k_y-\sigma_y k_z)$ where $\sigma_i$ is the corresponding Pauli matrix. In GaAs/AlAs heterostructures $\alpha$ is of the order of $~$3.9$\times 10^{-12}$ eV-m\cite{DattaDas}. For an externally applied electric field of $10^6$Volt/m, the material Rashba term is a factor of $10^8$ larger than the bare Pauli spin-orbit strength at the same applied electric field.

Summarizing, we have established that a vector potential reformulation of the Pauli Hamiltonian leads to a gauge symmetry broken Yang-Mills type theory. Such symmetry breaking renders the corresponding gauge field a real observable field related to the intrinsic or external electric fields determining the spin-orbit strength. This formulation of the theory permits correctly accounting for the interplay between coupled matter and radiation fields, both with angular momentum content\cite{ZhangYangMills} and the correct formulation for the currents. The additional matter spin current contribution from the GSB term exactly cancels the London contribution, giving zero matter spin current. There is then only a nonzero contribution to the spin current from the radiative terms. We then demonstrate that the commutation of the translation operators with the $SU(2)$ generators on a periodic lattice, results in a quantization condition for spatially constant electric fields, only when such field is perpendicular to the plane.  
Recent proposals have addressed the measurement of spin currents\cite{SoninTinkham}, resorting to deformations and measurement of charge currents. Such measurement could shed light on the present GSB picture.   

\acknowledgments
We acknowledge fruitful discussions with Claudio Chamon and Prashant Sharma. EM acknowledges the hospitality of the Physics Department of Boston University where this work was initiated. This work was supported by CNRS-Fonacyt grant PI-2008000272.

\end{document}